\documentclass[aps,twocolumn,preprintnumbers,amsmath,amssymb,prb,superscriptaddress]{revtex4}
\usepackage[dvips]{epsfig}
\usepackage{subfigure}
\usepackage{graphicx}
\usepackage{longtable}
\usepackage[T1]{fontenc}
\usepackage[ansinew]{inputenc}
\usepackage{ulem}
\usepackage{color}
\usepackage{float}

\begin{document}

\title{Modification of the charge and magnetic order of a low dimensional ferromagnet by molecule-surface bonding}

\author{Johannes Seidel}
\email[]{jseidel@rhrk.uni-kl.de}
\affiliation{Department of Physics and Research Center OPTIMAS, University of Kaiserslautern, Erwin-Schroedinger-Strasse 46, 67663 Kaiserslautern, Germany}
\author{Eva S. Walther}
\affiliation{Department of Physics and Research Center OPTIMAS, University of Kaiserslautern, Erwin-Schroedinger-Strasse 46, 67663 Kaiserslautern, Germany}
\author{Sina Mousavion}
\affiliation{Department of Physics and Research Center OPTIMAS, University of Kaiserslautern, Erwin-Schroedinger-Strasse 46, 67663 Kaiserslautern, Germany}

\author{Dominik Jungkenn}
\affiliation{Department of Physics and Research Center OPTIMAS, University of Kaiserslautern, Erwin-Schroedinger-Strasse 46, 67663 Kaiserslautern, Germany}
\author{Markus Franke}
\affiliation{Peter Gr\"unberg Institut (PGI-3), Forschungszentrum J\"ulich, 52425 J\"ulich, Germany}
\affiliation{J\"ulich-Aachen Research Alliance (JARA) -- Fundamentals of Future Information Technology, 52425 J\"ulich, Germany}
\author{Leah L. Kelly}
\affiliation{Department of Physics and Research Center OPTIMAS, University of Kaiserslautern, Erwin-Schroedinger-Strasse 46, 67663 Kaiserslautern, Germany}
\author{Ahmed Alhassanat}
\affiliation{Johannes-Gutenberg-Universit\"at Mainz, Institut f\"ur Physik, 55128 Mainz, Germany}
\author{Hans-Joachim Elmers}
\affiliation{Johannes-Gutenberg-Universit\"at Mainz, Institut f\"ur Physik, 55128 Mainz, Germany}
\affiliation{Graduate School of Excellence Materials Science in Mainz, Erwin Schroedinger Strasse 46, 67663 Kaiserslautern, Germany}
\author{Mirko Cinchetti}
\affiliation{Experimentelle Physik VI, Technische Universit\"at Dortmund, 44221 Dortmund, Germany}
\author{Christian Kumpf}
\affiliation{Peter Gr\"unberg Institut (PGI-3), Forschungszentrum J\"ulich, 52425 J\"ulich, Germany}
\author{Martin Aeschlimann}
\affiliation{Department of Physics and Research Center OPTIMAS, University of Kaiserslautern, Erwin-Schroedinger-Strasse 46, 67663 Kaiserslautern, Germany}
\author{Benjamin Stadtm\"uller}
\affiliation{Department of Physics and Research Center OPTIMAS, University of Kaiserslautern, Erwin-Schroedinger-Strasse 46, 67663 Kaiserslautern, Germany}
\affiliation{Graduate School of Excellence Materials Science in Mainz, Erwin Schroedinger Strasse 46, 67663 Kaiserslautern, Germany}

\date{\today}
\begin{abstract}
The ability to design and control the spin and charge order of low dimensional materials on the molecular scale offers an intriguing pathway towards the miniaturization of spintronic technology towards the nanometer scale. In this work, we focus on the adsorption induced modifications of the magnetic and electronic properties of a low dimensional ferromagnetic surface alloy after the adsorption of the prototypical organic molecule perylene-3,4,9,10-tetracarboxylic dianhydride (PTCDA). For this metal-organic interface, we observe the formation of a localized $\sigma$-like bond between the functional molecular groups and the surface alloy atoms. This strong chemical bonding coincides with a lifting of the characteristic surface alloy band structure and a reduction of the magnitude of the local magnetic moments of the Dy atoms by $18\,$\%. We attribute both findings to a mixing of spin-degenerate molecular states with spin-split states of the Dy-Ag surface alloy via the $\sigma$-like bonds between PTCDA and the Dy surface alloy atoms. Our findings clearly demonstrate the potential of tailored molecule-surface $\sigma$-bonds to control not only the electronic but also the magnetic order of low dimensional materials.
 \end{abstract}

\maketitle

\section*{Introduction}
One of the great challenges for next generation information technology is to develop novel concepts to control charge and spin carriers on ever-smaller length scales\cite{Yu2014,Kang2005,Kelley2004}. So far, most approaches for the miniaturization of spintronic assemblies rely on the continuous advances of nanofabrication methods which eventually resulted in the realization of prototypical spintronic structures with sub-$\mu$m sizes\cite{Mailly2009,Biswas2012,Stepanova2012}. 
Recently, an alternative route towards nanoscale spintronic units has emerged by the discovery of low dimensional magnetic systems. The most recent member of this family are 2D magnetic van der Waals materials such as CrI$_3$ or VSe$_2$\cite{Huang2018,Gibertini2019,Yu2019}. These 2D materials and their heterostructures exhibit a variety of intrinsic spin functionalities and are inherently nanoscale materials \cite{Gibertini2019,Tian2016,Intemann2015,Xing2017}. However, despite these great opportunities, they are still challenging to produce with sufficient quality and are extremely difficult to manipulate, for instance by chemical functionalization. \\
This is clearly different for spin-textured 2D surface alloys. These highly flexible low dimensional materials typically consist of two noble metal atoms and one heavy metal alloy atom per surface unit cell. From a spintronic point of view, they are particular intriguing due to their surface band structure with a Rashba-type spin split hybrid surface state that is responsible for the spin functionalities of these 2D systems. Crucially, there are clear pathways to chemically functionalize the spin-spit surface state of these surface alloys by the adsorption of organic molecules \cite{BSPRB2016,BSPRL2016,Friedrich.2017,Cinchetti2017}. For instance, we recently demonstrated that the surface band structure of heavy metal noble metal surface alloys can only be altered by tailored $\sigma$-bonds formed between functional groups of the molecular adsorbates and the surface alloy atoms \cite{BSPRB2016,BSPRL2016}. Theoretical predictions supported these experimental findings by showing that the spin splitting of the surface state can be controlled by tailored molecule-surface bonds \cite{Friedrich.2017}. Apart from spin-textured materials, theoretical predictions have also provided first indications for the profound influence of tailored molecule-surface bonds for the magnetic surface order of bulk materials or thin films, for instance, by a magnetic hardening or softening of the underlying metal atoms depending on the type and strength of the molecule-surface bond\cite{skyhook1,skyhooks1,skyhooks2}. \\

In this Letter, we take the next crucial step towards the functionalization of magnetic order in low dimensional systems by exploring chemical control schemes for 2D magnetic surface alloys by tailored molecule-surface bonds. 2D ferromagnetic surface alloys typically consist of rare earth atoms in a noble metal host matrix and exhibit a ferromagnetic order at low sample temperature\cite{Ormaza.2016, Seidel2019}. This magnetic order is caused by an indirect RKKY-like exchange coupling between the localized 4f electrons, mediated by spin-polarized valence electrons of a rare earth-noble metal hybrid surface state. This potentially opens two possibilities for the manipulation of the magnetic order: (i) direct control of the magnetic order by local manipulation of the magnetic moments of the rare earth atoms or (ii) indirect control of the magnetic order by altering the spin-polarized electrons of the hybrid surface state.  \\
Here, we demonstrate the adsorption induced modification of the magnetic and charge order of the 2D ferromagnetic surface alloy DyAg$_2$ \cite{Seidel2019} by the adsorption of a selected organic molecule, namely perylene-3,4,9,10-tetracarboxylic dianhydride (PTCDA). This molecule can show two fundamentally different molecule-surface interactions, either by the preferential formation of delocalized $\pi$-bonds between the aromatic molecular system and the surface alloy or by localized $\sigma$-bonds between functional molecular oxygen groups and the surface alloy atoms. In our work, the bonding type and strength across the interface is characterized by its chemical and geometric signatures using core level photoemission spectroscopy (XPS) and the normal incident x-ray standing waves (NIXSW) technique. The latter method allows us to determine the vertical position of each atomic species with high accuracy ($<0.04\,$\AA) and chemical sensitivity.  The corresponding changes in the charge and magnetic order for both cases are investigated using momentum-resolved photoemission spectroscopy and magnetic x-ray circular dichroism (XMCD).  

\section*{Results}

\begin{figure}
	\centering
	\includegraphics[width=60mm]{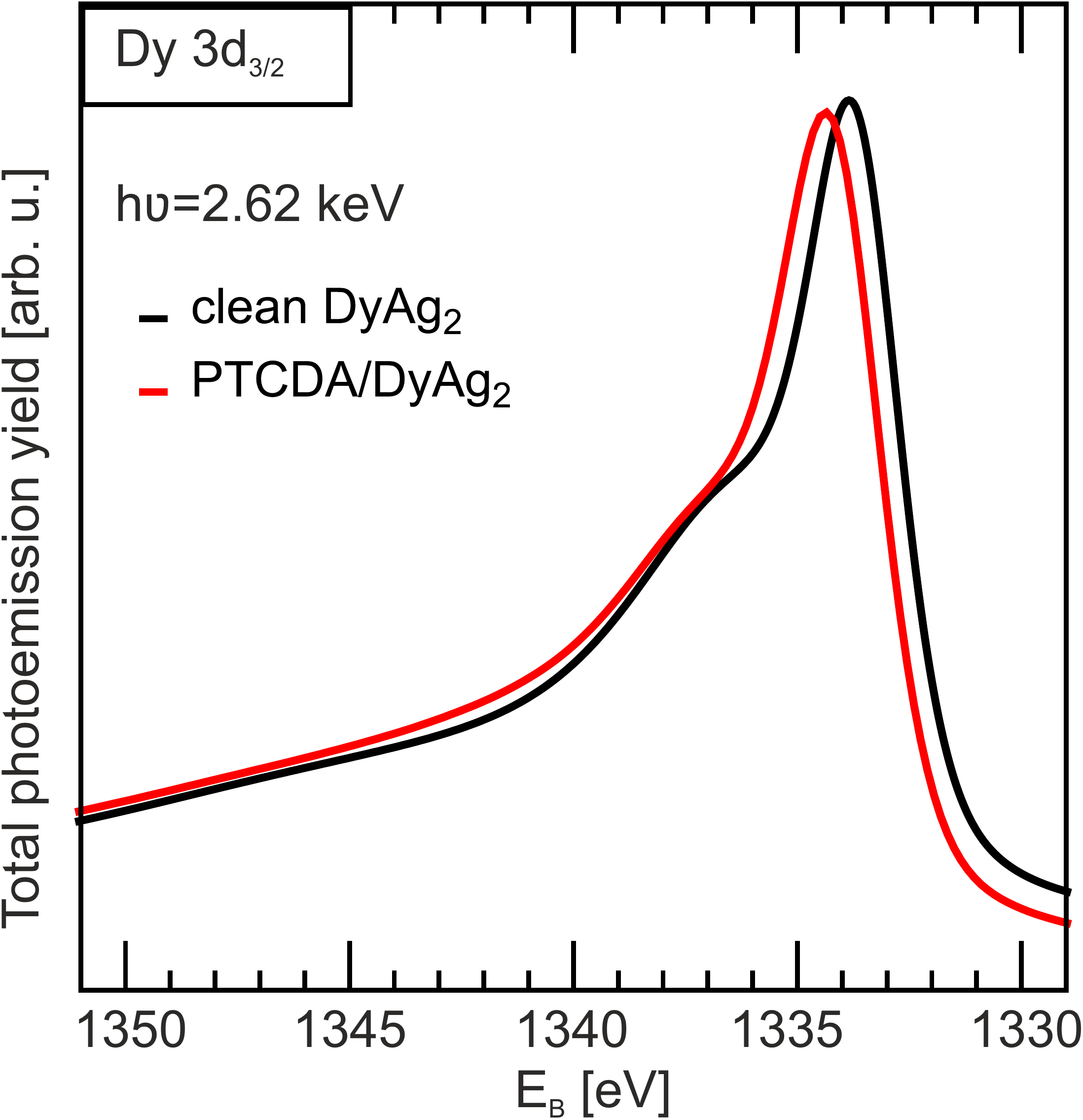} 
	\caption{XPS data of the core level signature of the Dy 3d$_{3/2}\,$core level for the clean alloy (black) and after the adsorption of PTCDA (red). The spectra were obtained with photon energies of h$\nu$=2.62 keV, in grazing emission geometry. }
\label{Fig:Fig1}
\end{figure}

The structural, electronic and magnetic properties of the DyAg$_2$ surface alloy were investigated prior and after the adsorption of PTCDA. The bare DyAg$_2$ surface alloy exhibits a long range ordered $\sqrt{3}\times \sqrt{3}R30^{\circ}$-like superstructure with an additional large scale Moir\'e pattern that can be characterized by a $16 \times 16\,$ superstructure. The lateral order of the surface alloy is not altered by the adsorption of a monolayer film of PTCDA on DyAg$_2$. The molecular film is completely disordered as indicated by the absence of any additional diffraction signatures after the formation of the metal-organic hybrid system (more information about the sample preparation can be found in the supporting information). 

We start our discussion with the characterization of the bonding type and strength at the PTCDA/DyAg$_2$ interface. The chemical environment of all atomic species is reflected in the lineshape and binding energy of the corresponding core level signatures. The Dy 3d$_{3/2}$ core level of the bare as well as of the PTCDA covered DyAg$_2$ surface alloy are shown in Fig.~\ref{Fig:Fig1}. The core level signature of the bare DyAg$_2$ surface alloy (black graph) reveals an asymmetric lineshape with a broad tail towards higher binding energy. The main peak is located at E$_{B}$=1333.8 $\pm 0.4\,$eV  which is $0.8\,$eV larger compared to its position in the bulk \cite{wagner1979handbook}. After the adsorption of PTCDA, the binding energy of the Dy 3d$_{3/2}$ main line (red curve) increases by $0.5\,$eV. These observations directly point to a significant charge redistribution and a strong chemical interaction between the PTCDA layer and the Dy surface alloy atoms alloy. \\

\begin{figure}
	\centering
	\includegraphics[width=85mm]{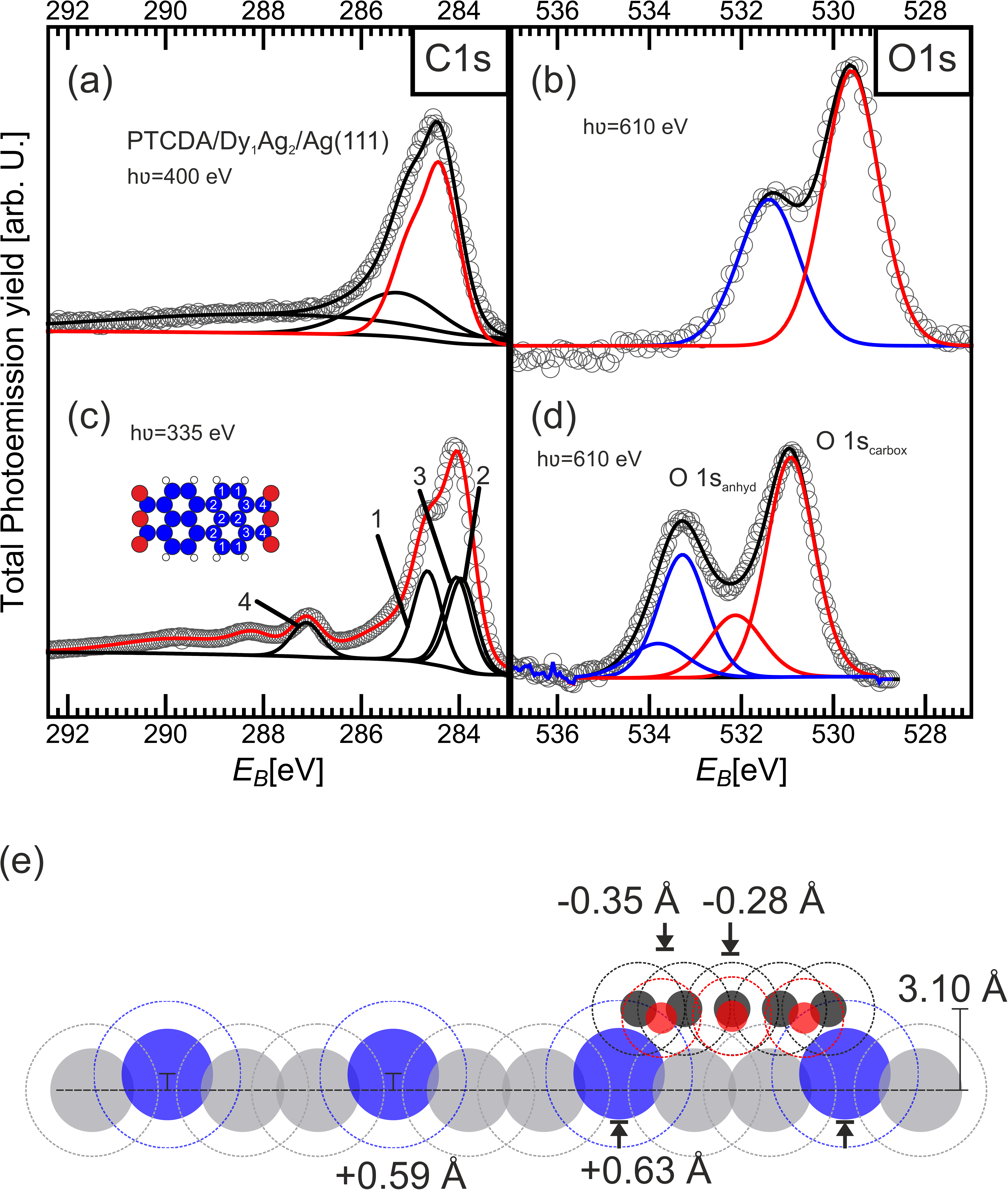} 

		\caption{(a): XPS spectra for the C 1s and (b) O 1s core levels of PTCDA for adsorption on DyAg$_2$. (c) C 1s and (d) O 1s core levels of PTCDA for the adsorption on Ag(111). (e) True scale vertical adsorption configuration of PTCDA on DyAg$_2$ as determined by NIXSW.}
\label{Fig:Fig2}
\end{figure}

The molecular core level signatures of the PTCDA/DyAg$_2$ interface are summarized in Fig.~\ref{Fig:Fig2}. 
The core level signatures of the carbon backbone and the oxygen end groups of PTCDA/DyAg$_2$ are shown in Fig.~\ref{Fig:Fig2}(a),(b). The ones for PTCDA/Ag(111) are included in Fig.~\ref{Fig:Fig2}(c),(d) as a reference \cite{BSPRB2016}. We observe a broadening of the C 1s lineshape and a small shift of E$_{B}=0.3\pm 0.1 \,$eV towards larger binding energies for PTCDA on the surface alloy compared to the bare Ag(111). The most peculiar difference between the adsorption of PTCDA on both surfaces is the large energy shift of $\Delta$E$_{B}=2.1\pm 0.1\,$eV of the carboxylic (marked as O 1s$_{carbox}$) and the anhydride (marked as O 1s$_{anhyd}$) O 1s core level signature to smaller binding energies, shown in panels (b) and (d). On the one hand, this large shift can explain the absence of the typically energetically well separated C-O carbon signature in the C 1s core level spectra of the PTCDA/DyAg$_2$ interface (marked as species 4 in Fig.~\ref{Fig:Fig2}(c) in the C~1s core signature of PTCDA/Ag(111) ). On the other hand, the rather large energetic shifts of the Dy 3d$_{3/2}$ and to O~1s species of the surface alloy and the PTCDA molecule are a first strong indication for the formation of a local $\sigma$-like bond between the PTCDA oxygen end groups and the Dy surface alloy atoms.    \\
The formation of such a local Dy-O bond is further supported by the vertical adsorption geometry of the PTCDA/DyAg$_2$ interface which is determined using the NIXSW method. This method allows us to determine the vertical adsorption position D$^H$ of each chemically different atomic species above the surface plane of the Ag(111) crystal together with the so called coherent fraction F$^H$. The latter describes the degree of vertical order of an atomic species above the surface with F$^H=1$ corresponding to perfect vertical ordering and F$^H=0$ to a completely random vertical distribution of the respective atoms (for more information, see supporting information). The experimental results of the NIXSW experiment are summarized in Tab.~\ref{Tab:Tab1}, the corresponding vertical adsorption geometry of the organic molecules is illustrated in Fig.~\ref{Fig:Fig2}(e). For the bare surface alloy, the Dy atoms are located $0.59 \pm0.02\,$\AA$\,$ above the plane of the Ag surface atoms (black horizontal line in Fig.~\ref{Fig:Fig2}(e) ) which is comparable to the vertical relaxation of the heavy metals Pb and Bi in their surface alloy structure on Ag(111) \cite{BSPRB2016,Gierz.2010}. After the adsorption of PTCDA, we find a small vertical displacement of the Dy alloy atoms by $\Delta z= 0.04\pm 0.03\,$\AA, i.e., the Dy atoms are located at D$^H=0.63\pm0.02\,$\AA. 

Turning to the vertical adsorption geometry of the atomic species of PTCDA, we find the carbon backbone  at a vertical position of D$^H= 3.10\pm0.01\,$\AA$\,$ above the topmost Ag(111) plane. Both oxygen end groups are located $0.30\,$\AA$\,$ below the molecular backbone, i.e., both oxygen atoms bend towards the atoms of the DyAg$_2$ surface alloy. This is fully in line with the strong chemical shifts observed in the Dy and O core level data and hence points to the formation of a local $\sigma$-like bond between the oxygen end groups of PTCDA and the Dy surface alloy atoms. This is also highlighted by the empiric radii of the atomic species (solid circles) and the elemental van-der-Waals radii (dashed circles) in the vertical adsorption model in Fig.~\ref{Fig:Fig2}(e). We observe a significant overlap of the van der Waals radii of the O atoms and the Dy alloy atoms further indicates a chemical interaction between both atoms. To quantify the bonding strength, we calculated the bonding distances between the atomic species of the molecule and the DyAg$_2$ surface alloy and normalized these distances to the sum of the van der Waals radii of the corresponding elements:
\begin{equation}
d^{N}_{A-B}=\frac{d_{A-B}}{d_{VdW}^{A}+d_{VdW}^{B}} .
\end{equation}
This normalized distance can be understood as a measure for the bonding strength between two atoms A and B for which $d^{N}_{A-B}\approx 1\,$ indicates a non-chemical van der Waals like interaction while $d^{N}_{A-B}<1$ suggest an at least partial chemical interaction due to the overlap of electronic wave functions of both atomic species. 
For the case of PTCDA/DyAg$_2$, we obtain normalized bonding distances for the carbon backbone as well as for both types of O atoms of $d^{N}_{C-Dy}=0.54\pm 0.004$, $d^{N}_{O_{carb  }-Dy}=0.49\pm 0.01 $, and $d^{N}_{O_{anhyd}-Dy}=0.51\pm 0.02 $. These normalized distances are significantly smaller than the corresponding bonding distances for  the PTCDA/Ag(111) interface ($d^{N}_{C-Ag}=0.68$, $d^{N}_{O_{carb}-Ag}=0.68$, and $d^{N}_{O_{anhyd}-Ag}=0.71$). This suggests the formation of even stronger $\sigma$-bonds between PTCDA and the DyAg$_2$ surface alloy compared to PTCDA/Ag(111). In addition, the small normalized bonding distance between the carbon backbone and the surface alloy atoms even suggests a large overlap between the delocalized $\pi$ orbitals of the carbon backbone and the surface states.\\
\begin{table*}[t]
\begin{center}
\caption{NIXSW fitting results and corresponding adsorption heights of all chemically inequivalent species of the PTCDA/DyAg$_2$ interface. For comparison, adsorption heights on the Ag(111) surface are included. We used non-dipolar correction parameters of $\gamma$=1.06219 (0.9757) for the C1s (O1s) emission lines, respectively. The experimental geometry is given by the Bragg angle of $\theta$=86.5$^{\circ}$ (3.5$^{\circ}\,$ off normal incidence) and by $\phi$=75$^{\circ}$ (photoelectron emission angle relative to incident beam). For more details see\cite{VANSTRAATEN2018106} and \cite{toricelli} for instance. For the analysis of the Dy 3d core level, we only considered the photoemission yield in a narrow angle range close to an emission angle of $90^\circ$ to minimize the influence of non-dipolar contributions to the photoemission signal.
Reference data of PTCDA on Ag(111) from \cite{Hauschild2010}.}
\begin{tabular}{ p{0.16\textwidth} p{0.16\textwidth} p{0.16\textwidth} p{0.16\textwidth} p{0.16\textwidth} p{0.16\textwidth} } 
\hline \hline
 &$F^H$ &   $D^{H}_{\mathrm{DyAg}}\,$[\AA] & $d^{N}_{X-Dy} $&   $D^{H}_{\mathrm{Ag}}\,$[\AA]  \\
\hline 
Dy 3d$_{\mathrm{clean (PTCDA)}}$& $0.76 \pm 0.08$   & $0.59 \pm 0.02$ & - & -   \\
Dy 3d$_{\mathrm{PTCDA}}$& $0.74 \pm 0.03$ & $0.63 \pm 0.02$ & - & -  \\
C1s$_{\mathrm{PTCDA}}$& $0.30 \pm 0.03$& $3.10 \pm 0.01$ & $0.54\pm 0.004$ &  $2.86 \pm 0.01$  \\
O1s$_{\mathrm{anhydrid}}$& $0.36 \pm 0.04$   & $2.82 \pm 0.03$ & $0.51\pm 0.02$& $2.98\pm0.08$ \\
O1s$_{\mathrm{carbox}}$& $0.32 \pm 0.03$ &$2.75 \pm 0.01$ & $0.49\pm 0.01$ &  $2.66\pm0.02$ \\

\hline \hline
\end{tabular}
\label{Tab:Tab1}
\end{center}
\end{table*}
We propose that the strong molecule surface interaction at the PTCDA/DyAg$_2$ interface is responsible for the low coherent fractions observed for all molecular species in our NIXSW analysis (see Tab.~\ref{Tab:Tab1}). The coherent fractions F$^H$ of the molecular species in our model system are smaller compared to typical values reported for highly ordered molecular films of planar molecules in a flat adsorption geometry on surfaces. Such low coherent fractions are typically associated with a significant vertical disorder and could be attributed to a poor quality of the molecular film. However, in our case, we propose that the low coherent fraction is a direct result of the strong molecule-surface interaction, which prevents the formation of any long-range molecular order and thus the existence of well-defined adsorption sites. This could result in the existence of multi adsorptions sites, for instance, on top of the Dy atoms or the Ag atoms which could lead not only to flat adsorption geometries of PTCDA, but also to tilted molecular configurations or arch-like distortions of the molecular backbone as reported for PTCDA on bare and modified Ag(110) surface\cite{Mercurio2013}.


\begin{figure}
	\centering
	\includegraphics[width=85mm]{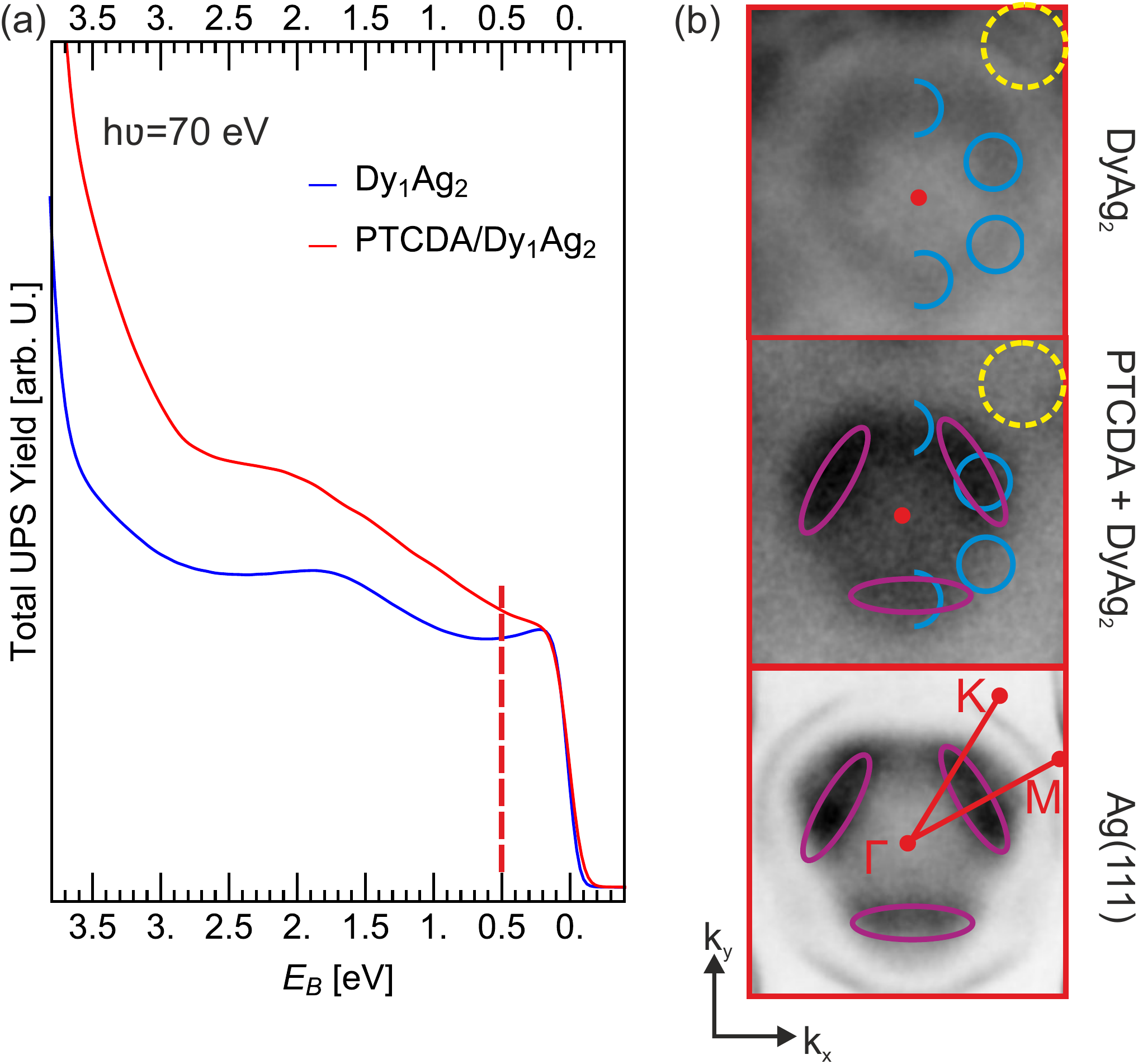} 

		\caption{(a): Total UPS yield of the electronic valence structure obtained by momentum microscopy. Blue graph: clean DyAg$_2$ surface alloy, red graph: PTCDA/DyAg$_2$. (b) Constant energy maps at E$_B$=0.5 eV for DyAg$_2$ (top), PTCDA/DyAg$_2$ (middle) and Ag(111) (bottom).}
\label{Fig:Fig3}
\end{figure}

We now focus on the electronic structure of the PTCDA/DyAg$_2$ interface investigated by momentum-resolved photoemission spectroscopy. The momentum integrated and momentum-resolved photoemission data are presented in Fig.~\ref{Fig:Fig3}. We start with the total photoemission yield for the bare DyAg$_2$ surface alloy which is shown as blue solid line in Fig.~\ref{Fig:Fig3}(a). The two distinct peaks in the valence band region at E$_F$ and at E$_B$=1.8 eV can be attributed to the Dy 4f states\cite{Seidel2019}. After the adsorption of PTCDA, we find an almost featureless total photoemission yield. The signal of the Dy 4f states is significantly attenuated and is replaced by a spectral broad feature located in the energy region between $1.0\,$ eV and $2.5\,$eV. Such a featureless valence band structure of PTCDA is very surprising compared to recent findings on noble metal surfaces\cite{Tautz2007,Kawabe2008,Romaner2009} and heavy-metal noble metal surface alloys\cite{Cottin2014,BSPRL2016,SeidelSnAg2019}.The absence of any-well-defined molecular features in the valence band structure is more commonly observed for the adsorption of planar or small organic molecules on transition metal surfaces  \cite{Tiba2006, Droghetti.2016}. We therefore attribute the very broad molecular spectral density to the extremely strong chemical interaction between the $\pi$-orbitals of PTCDA and the Dy-Ag surface alloy states.

To uncover adsorption-induced modifications of the surface alloy band structure, we turn to the momentum resolved photoemission pattern around the center of the surface Brillouin zone (SBZ) of the DyAg$_2$ surface alloy in Fig.~\ref{Fig:Fig3}(b). In this section of the SBZ, the band structure of the DyAg$_2$ surface alloy (top panel) reveals a sixfold symmetric emission pattern with six well defined emission maxima arranged in a circle around the $\bar{\Gamma}$-point (blue circles). In addition, the circular emission pattern of the Dy-Ag hybrid surface state is visible at the $\bar{\Gamma}$-point of the next SBZ, highlighted by a yellow circle. This hybrid surface state \cite{Seidel2019} has been identified as the mediator for the ferromagnetic order of the Dy 4f electrons of the DyAg$_2$ surface alloy.

After the adsorption of PTCDA, the emission pattern of the DyAg$_2$ surface band structure changes significantly, see middle panel of Fig.~\ref{Fig:Fig3}(b). The six emission maxima surrounding the $\bar{\Gamma}$-point vanish and are replaced by a threefold symmetric emission pattern with elongated emission features in momentum space, highlighted by purple ellipsoids. The elongated emission feature of the PTCDA/DyAg$_2$ interface are almost identical to the emission pattern of the bare Ag(111) surface which has been included as a reference in the bottom panel of Fig.~\ref{Fig:Fig3}(b). More importantly, the Dy-Ag hybrid surface state vanishes completely after the adsorption of PTCDA on the surface alloy. In this way, the surface band structure of the metallic side of the PTCDA/DyAg$_2$ interface resembles the surface band structure of the bare Ag(111) surface. More details of the spectroscopic changes of the DyAg$_2$ surface alloy are reported in the supplementary information. 
We attribute the adsorption-induced changes in the DyAg$_2$ surface band structure to the strong chemical bonding between the PTCDA molecules and the surface alloy atoms. In particular, the direct $\sigma$-like bonding between the anhydride and carboxylic oxygen endgroups of PTCDA and the Dy surface alloy atoms results in a clear charge transfer across the interface which could severely alter the electronic properties of the Dy-derived surface states, i.e., of the Dy-Ag hybrid surface state.  

\begin{figure}
\centering
\includegraphics[width=85mm]{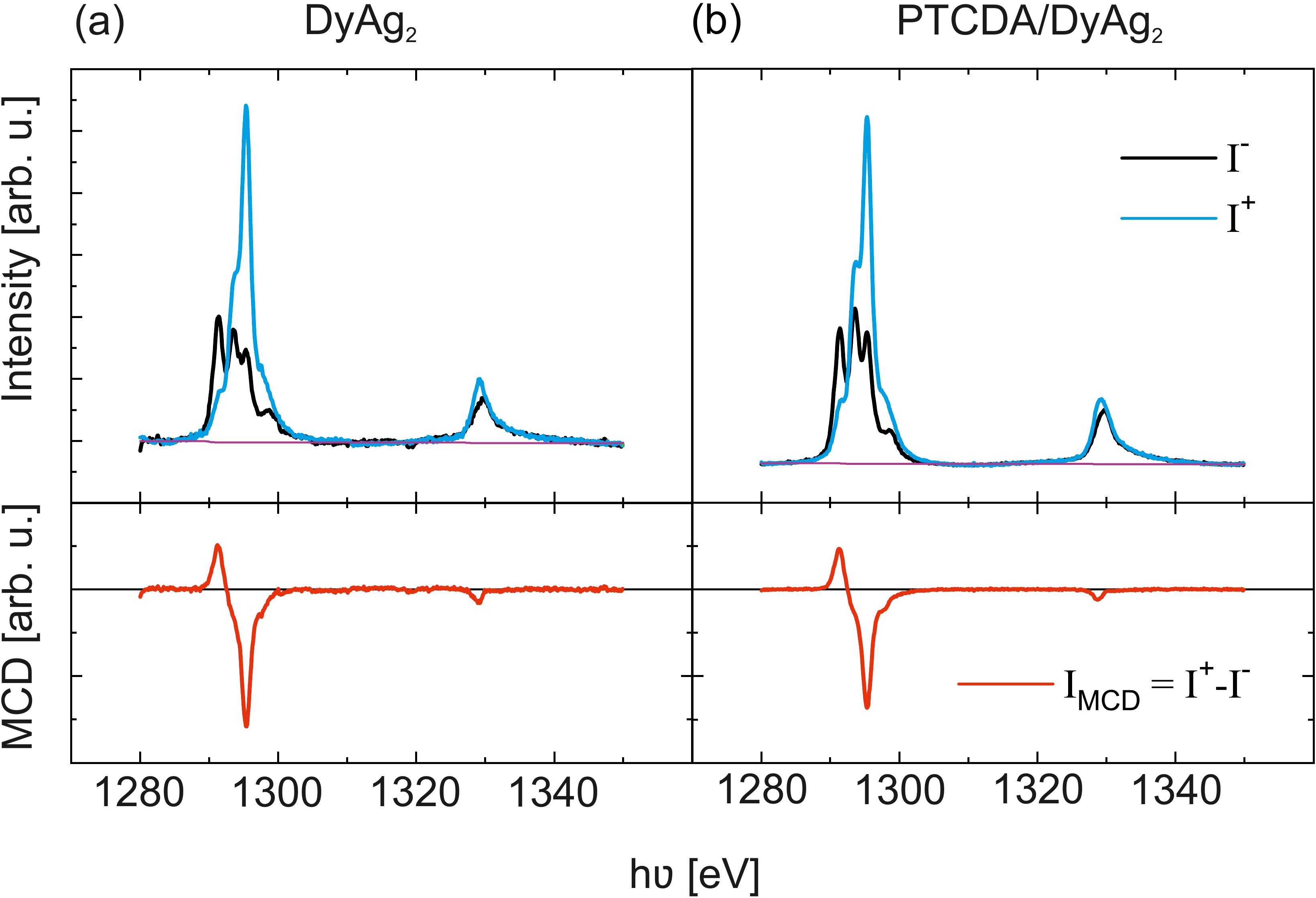}
\caption{XMCD traces recorded for in plane magnetization for the clean DyAg$_2$ surface alloy and upon adsorption of PTCDA.}
\label{Fig:Fig5}
\end{figure}

Changes in the magnetic or spin-dependent properties of materials typically result from modifications of their electronic and geometric structure \cite{Mirhosseini2010,Meier2009,Pacil2006}. This is also the case for the PTCDA/DyAg$_2$ surface alloy as demonstrated by our XMCD measurements. We recorded the XMCD signal of Dy for the DyAg$_2$ surface alloy prior and after the adsorption of PTCDA in a magnetic field fields of $\pm 7\,$T at a sample temperature of $T=10\,$K. The total absorption yield curves of the Dy 3$d\rightarrow4f\,$ transitions (M$_{4,5}$-edge) are shown in the top panels of Fig.~\ref{Fig:Fig5}(a) for the clean and (b) the PTCDA covered surface alloy, the corresponding XMCD traces are shown below. Both XMCD signals reveal the typical sign change at both Dy absorption edges . Using the XMCD sum-rule analysis introduced in Ref.\cite{Thole1985},we can extract the total magnetic moment of the localized Dy atoms prior and after the adsorption of PTCDA. We find a total magnetic moment of $\mu_{mag}=4.5\mu_B$ for the bare surface alloy, which decrease significantly to $\mu_{mag}=3.71\mu_B$   upon the adsorption of PTCDA. This reduction of the local magnetic moments at the Dy sites suggests a modification of the occupation of the Dy 4f electrons due to the adsorption of PTCDA. This observation is consistent with the direction of charge redistribution between the Dy atoms and the O end groups of PTCDA coinciding with the formation of a local $\sigma$-like bond. Since the charge distribution of PTCDA is spin-degenerate, the mixing of the spin-degenerate states of PTCDA with the spin-split states of Dy results in an average reduction of the spin-polarization of the electronic states of the Dy surface atoms. We hence propose that this mixing of spin degenerate molecular and spin-split metallic states results in a reduction of the local magnetic moments on the Dy sites.

\section*{Conclusion}
In this work, we demonstrated a new pathway to modify the magnetic properties of low dimensional magnetic systems by adsorption of organic molecules. Using a DyAg$_2$ surface alloy as model system, we uncovered a clear change in the magnetic moments of the surface atoms by the formation of local $\sigma$-like bonds between functional molecular groups and surface alloy atoms. 
First, we find clear signatures for a chemical bonding between the prototypical molecule PTCDA and the ferromagnetic surface alloy DyAg$_2$ in the chemical environment and the vertical adsorption configuration of the metal-organic interface. Our core level spectroscopy reveals an interaction induced charge redistribution between the Dy surface alloy atoms and the functional oxygen end groups of PTCDA. These findings are consistent with an overall small bonding distance of $2.47\pm 0.02\,$\AA\ between the PTCDA carbon backbone and the surface alloy atoms as observed by the NIXSW technique. More importantly, we find an bending of the PTCDA oxygen endgroups towards the surface alloy atoms coinciding with an additional vertical relaxation of the Dy surface alloy atoms towards the molecular layer. These experimental findings clearly point to the formation of (i) a delocalized $\pi$-bonding between the aromatic molecular orbitals of the carbon backbone and the delocalized surface electrons as well as of (ii) a local $\sigma$-like bonding between the oxygen endgorups of PTCDA and the Dy surface alloy atoms. The O-Dy bonding distance is significantly smaller than the corresponding O-Pb distance for PTCDA/PbAg$_2$\cite{BSPRL2016,BSPRB2016}, which points to an even stronger local coupling between PTCDA and the DyAg$_2$ surface alloy. 
The strong chemical interaction at the PTCDA/DyAg$_2$ interface is also directly reflected in the electronic and magnetic properties of the interface. We observed a metallic spectral density of the molecular film resembling the adsorption of aromatic molecules on transition metal surfaces. In addition, the characteristic surface band structure of the DyAg$_2$ surface alloy is completely suppressed by the adsorption of PTCDA. Similarly, we observe a reduction of the local magnetic moment of the Dy atoms by $18\,$\%.
These changes in the electronic and magnetic surface properties can only be attributed to the formation of local, $\sigma$-like bonds between the organic molecules and the surface alloy. Such bonds usually coincide with a strong hybridization of molecular and metallic states of the ferromagnetic surface with different orbital character and spin degeneracy. We propose that this strong hybridization of molecular and metallic states is responsible for the modification of the orbital character of the Dy-Ag surface state that eventually results in a lifting of its hole-like band dispersion. In a similar way, the mixing of spin-degenerate and spin-split states of the molecular and metallic side of the interface results in a reduction of imbalance of the spin-polarized Dy 4f electrons of the surface alloy and hence in a reduction of the local magnetic moments on the Dy sites.

In conclusion, we have demonstrated the crucial role of molecule-surface bonding on the magnetic order of ferromagnetic binary surface alloys. Our findings present a clear pathway for controlling the magnetic and electronic properties of a larger class of low dimensional materials. This will pave the way towards controlling low dimensional materials on the intrinsic nanometer scale of molecular complexes by the formation of tailored molecule surface bonds. 

\section*{Acknowledgements}
This research was funded by the Deutsche Forschungsgemeinschaft (DFG, German Research Foundation) - TRR 173 - 268565370 (Project A09 and B05). B.S. and H.J.E. thankfully acknowledge financial support from the Graduate School of Excellence Mainz (Excellence initiative DFG/GSC 266), M.F. and C.K. were supported by DFG via SFB 1083. L.L.K. thankfully acknowledges financial support from the Carl Zeiss Stiftung. M.C. acknowledges funding from the European Research Council (ERC) under the European Union's Horizon 2020 research and innovation programme (Grant Agreement No. 725767-hyControl). We thank Diamond Light Source for access to beamline I09 (through proposal SI16086-1). We are very grateful for the support by the beamline staff during the experiment, in particular by P. K. Thakur, D. A. Duncan, T.-L. Lee, and D. McCue. We thank HZB for the allocation of synchrotron radiation beamtime.
We thank elettra Sincrotrone Trieste for the allocation of synchrotron radiation beamtime at the NanoESCA endstation. We are very grateful for the support by the beamline staff during the experiment, in particular by V. Feyer, G. Zamborlini and M. Jugovac.
\providecommand{\latin}[1]{#1}
\makeatletter
\providecommand{\doi}
  {\begingroup\let\do\@makeother\dospecials
  \catcode`\{=1 \catcode`\}=2 \doi@aux}
\providecommand{\doi@aux}[1]{\endgroup\texttt{#1}}
\makeatother
\providecommand*\mcitethebibliography{\thebibliography}
\csname @ifundefined\endcsname{endmcitethebibliography}
  {\let\endmcitethebibliography\endthebibliography}{}


\begin{mcitethebibliography}{38}
\providecommand*\natexlab[1]{#1}
\providecommand*\mciteSetBstSublistMode[1]{}
\providecommand*\mciteSetBstMaxWidthForm[2]{}
\providecommand*\mciteBstWouldAddEndPuncttrue
  {\def\EndOfBibitem{\unskip.}}
\providecommand*\mciteBstWouldAddEndPunctfalse
  {\let\EndOfBibitem\relax}
\providecommand*\mciteSetBstMidEndSepPunct[3]{}
\providecommand*\mciteSetBstSublistLabelBeginEnd[3]{}
\providecommand*\EndOfBibitem{}
\mciteSetBstSublistMode{f}
\mciteSetBstMaxWidthForm{subitem}{(\alph{mcitesubitemcount})}
\mciteSetBstSublistLabelBeginEnd
  {\mcitemaxwidthsubitemform\space}
  {\relax}
  {\relax}

\bibitem[Yu \latin{et~al.}(2014)Yu, Pr{\'{e}}vot, and Sivula]{Yu2014}
Yu,~X.; Pr{\'{e}}vot,~M.~S.; Sivula,~K. Multiflake Thin Film Electronic Devices
  of Solution Processed 2D {MoS}2 Enabled by Sonopolymer Assisted Exfoliation
  and Surface Modification. \emph{Chem Mater} \textbf{2014}, \emph{26},
  5892--5899\relax
\mciteBstWouldAddEndPuncttrue
\mciteSetBstMidEndSepPunct{\mcitedefaultmidpunct}
{\mcitedefaultendpunct}{\mcitedefaultseppunct}\relax
\EndOfBibitem
\bibitem[Kang \latin{et~al.}(2005)Kang, Davidson, Wisitsora-at, Wong, Takalkar,
  Subramania, Kerns, and Hofmeister]{Kang2005}
Kang,~W.; Davidson,~J.; Wisitsora-at,~A.; Wong,~Y.; Takalkar,~R.;
  Subramania,~K.; Kerns,~D.; Hofmeister,~W. Diamond and carbon-derived vacuum
  micro- and nano-electronic devices. \emph{Diam Relat Mater} \textbf{2005},
  \emph{14}, 685--691\relax
\mciteBstWouldAddEndPuncttrue
\mciteSetBstMidEndSepPunct{\mcitedefaultmidpunct}
{\mcitedefaultendpunct}{\mcitedefaultseppunct}\relax
\EndOfBibitem
\bibitem[Kelley \latin{et~al.}(2004)Kelley, Baude, Gerlach, Ender, Muyres,
  Haase, Vogel, and Theiss]{Kelley2004}
Kelley,~T.~W.; Baude,~P.~F.; Gerlach,~C.; Ender,~D.~E.; Muyres,~D.;
  Haase,~M.~A.; Vogel,~D.~E.; Theiss,~S.~D. Recent Progress in Organic
  Electronics:~ Materials, Devices, and Processes. \emph{Chem Mater}
  \textbf{2004}, \emph{16}, 4413--4422\relax
\mciteBstWouldAddEndPuncttrue
\mciteSetBstMidEndSepPunct{\mcitedefaultmidpunct}
{\mcitedefaultendpunct}{\mcitedefaultseppunct}\relax
\EndOfBibitem
\bibitem[Mailly(2009)]{Mailly2009}
Mailly,~D. Nanofabrication techniques. \emph{Eur Phys J Spec Top}
  \textbf{2009}, \emph{172}, 333--342\relax
\mciteBstWouldAddEndPuncttrue
\mciteSetBstMidEndSepPunct{\mcitedefaultmidpunct}
{\mcitedefaultendpunct}{\mcitedefaultseppunct}\relax
\EndOfBibitem
\bibitem[Biswas \latin{et~al.}(2012)Biswas, Bayer, Biris, Wang, Dervishi, and
  Faupel]{Biswas2012}
Biswas,~A.; Bayer,~I.~S.; Biris,~A.~S.; Wang,~T.; Dervishi,~E.; Faupel,~F.
  Advances in top{\textendash}down and bottom{\textendash}up surface
  nanofabrication: Techniques, applications {\&} future prospects. \emph{Adv
  Colloid Interface Sci} \textbf{2012}, \emph{170}, 2--27\relax
\mciteBstWouldAddEndPuncttrue
\mciteSetBstMidEndSepPunct{\mcitedefaultmidpunct}
{\mcitedefaultendpunct}{\mcitedefaultseppunct}\relax
\EndOfBibitem
\bibitem[Stepanova and Dew(2012)Stepanova, and Dew]{Stepanova2012}
Stepanova,~M., Dew,~S., Eds. \emph{Nanofabrication}; Springer Vienna,
  2012\relax
\mciteBstWouldAddEndPuncttrue
\mciteSetBstMidEndSepPunct{\mcitedefaultmidpunct}
{\mcitedefaultendpunct}{\mcitedefaultseppunct}\relax
\EndOfBibitem
\bibitem[Huang \latin{et~al.}(2018)Huang, Clark, Klein, MacNeill,
  Navarro-Moratalla, Seyler, Wilson, McGuire, Cobden, Xiao, Yao,
  Jarillo-Herrero, and Xu]{Huang2018}
Huang,~B.; Clark,~G.; Klein,~D.~R.; MacNeill,~D.; Navarro-Moratalla,~E.;
  Seyler,~K.~L.; Wilson,~N.; McGuire,~M.~A.; Cobden,~D.~H.; Xiao,~D.; Yao,~W.;
  Jarillo-Herrero,~P.; Xu,~X. Electrical control of 2D magnetism in bilayer
  {CrI}3. \emph{Nat Nanotechnol} \textbf{2018}, \emph{13}, 544--548\relax
\mciteBstWouldAddEndPuncttrue
\mciteSetBstMidEndSepPunct{\mcitedefaultmidpunct}
{\mcitedefaultendpunct}{\mcitedefaultseppunct}\relax
\EndOfBibitem
\bibitem[Gibertini \latin{et~al.}(2019)Gibertini, Koperski, Morpurgo, and
  Novoselov]{Gibertini2019}
Gibertini,~M.; Koperski,~M.; Morpurgo,~A.~F.; Novoselov,~K.~S. Magnetic 2D
  materials and heterostructures. \emph{Nat Nanotechnol} \textbf{2019},
  \emph{14}, 408--419\relax
\mciteBstWouldAddEndPuncttrue
\mciteSetBstMidEndSepPunct{\mcitedefaultmidpunct}
{\mcitedefaultendpunct}{\mcitedefaultseppunct}\relax
\EndOfBibitem
\bibitem[Yu \latin{et~al.}(2019)Yu, Li, Herng, Wang, Zhao, Chi, Fu, Abdelwahab,
  Zhou, Dan, Chen, Chen, Li, Lu, Pennycook, Feng, Ding, and Loh]{Yu2019}
Yu,~W. \latin{et~al.}  Chemically Exfoliated VSe$_2$ Monolayers with
  Room-Temperature Ferromagnetism. \emph{Adv. Mater.} \textbf{2019}, \emph{31},
  1903779\relax
\mciteBstWouldAddEndPuncttrue
\mciteSetBstMidEndSepPunct{\mcitedefaultmidpunct}
{\mcitedefaultendpunct}{\mcitedefaultseppunct}\relax
\EndOfBibitem
\bibitem[Tian \latin{et~al.}(2016)Tian, Gray, Ji, Cava, and Burch]{Tian2016}
Tian,~Y.; Gray,~M.~J.; Ji,~H.; Cava,~R.~J.; Burch,~K.~S. Magneto-elastic
  coupling in a potential ferromagnetic 2D atomic crystal. \emph{2d Mater}
  \textbf{2016}, \emph{3}, 025035\relax
\mciteBstWouldAddEndPuncttrue
\mciteSetBstMidEndSepPunct{\mcitedefaultmidpunct}
{\mcitedefaultendpunct}{\mcitedefaultseppunct}\relax
\EndOfBibitem
\bibitem[Intemann \latin{et~al.}(2015)Intemann, Yao, Ding, Xu, Xin, Li, and
  Jen]{Intemann2015}
Intemann,~J.~J.; Yao,~K.; Ding,~F.; Xu,~Y.; Xin,~X.; Li,~X.; Jen,~A. K.-Y.
  Enhanced Performance of Organic Solar Cells with Increased End Group Dipole
  Moment in Indacenodithieno[3, 2-b]thiophene-Based Molecules. \emph{Adv Funct
  Mater} \textbf{2015}, \emph{25}, 4889--4897\relax
\mciteBstWouldAddEndPuncttrue
\mciteSetBstMidEndSepPunct{\mcitedefaultmidpunct}
{\mcitedefaultendpunct}{\mcitedefaultseppunct}\relax
\EndOfBibitem
\bibitem[Xing \latin{et~al.}(2017)Xing, Chen, Odenthal, Zhang, Yuan, Su, Song,
  Wang, Zhong, Jia, Xie, Li, and Han]{Xing2017}
Xing,~W.; Chen,~Y.; Odenthal,~P.~M.; Zhang,~X.; Yuan,~W.; Su,~T.; Song,~Q.;
  Wang,~T.; Zhong,~J.; Jia,~S.; Xie,~X.~C.; Li,~Y.; Han,~W. Electric field
  effect in multilayer Cr$_2$Ge$_2$Te$_6$: a ferromagnetic 2D material.
  \emph{2d Mater} \textbf{2017}, \emph{4}, 024009\relax
\mciteBstWouldAddEndPuncttrue
\mciteSetBstMidEndSepPunct{\mcitedefaultmidpunct}
{\mcitedefaultendpunct}{\mcitedefaultseppunct}\relax
\EndOfBibitem
\bibitem[Stadtm\"uller \latin{et~al.}(2016)Stadtm\"uller, Haag, Seidel, van
  Straaten, Franke, Kumpf, Cinchetti, and Aeschlimann]{BSPRB2016}
Stadtm\"uller,~B.; Haag,~N.; Seidel,~J.; van Straaten,~G.; Franke,~M.;
  Kumpf,~C.; Cinchetti,~M.; Aeschlimann,~M. Adsorption heights and bonding
  strength of organic molecules on a Pb-Ag surface alloy. \emph{Phys Rev B}
  \textbf{2016}, \emph{94}, 235436\relax
\mciteBstWouldAddEndPuncttrue
\mciteSetBstMidEndSepPunct{\mcitedefaultmidpunct}
{\mcitedefaultendpunct}{\mcitedefaultseppunct}\relax
\EndOfBibitem
\bibitem[Stadtm\"uller \latin{et~al.}(2016)Stadtm\"uller, Seidel, Haag, Grad,
  Tusche, van Straaten, Franke, Kirschner, Kumpf, Cinchetti, and
  Aeschlimann]{BSPRL2016}
Stadtm\"uller,~B.; Seidel,~J.; Haag,~N.; Grad,~L.; Tusche,~C.; van
  Straaten,~G.; Franke,~M.; Kirschner,~J.; Kumpf,~C.; Cinchetti,~M.;
  Aeschlimann,~M. Modifying the Surface of a Rashba-Split Pb-Ag Alloy Using
  Tailored Metal-Organic Bonds. \emph{Phys. Rev. Lett.} \textbf{2016},
  \emph{117}, 096805\relax
\mciteBstWouldAddEndPuncttrue
\mciteSetBstMidEndSepPunct{\mcitedefaultmidpunct}
{\mcitedefaultendpunct}{\mcitedefaultseppunct}\relax
\EndOfBibitem
\bibitem[Friedrich \latin{et~al.}(2017)Friedrich, Caciuc, Bihlmayer,
  Atodiresei, and Bl{\"u}gel]{Friedrich.2017}
Friedrich,~R.; Caciuc,~V.; Bihlmayer,~G.; Atodiresei,~N.; Bl{\"u}gel,~S.
  {Designing the Rashba spin texture by adsorption of inorganic molecules}.
  \emph{{New J Phys}} \textbf{2017}, \emph{19}, 043017\relax
\mciteBstWouldAddEndPuncttrue
\mciteSetBstMidEndSepPunct{\mcitedefaultmidpunct}
{\mcitedefaultendpunct}{\mcitedefaultseppunct}\relax
\EndOfBibitem
\bibitem[Cinchetti \latin{et~al.}(2017)Cinchetti, Dediu, and
  Hueso]{Cinchetti2017}
Cinchetti,~M.; Dediu,~V.~A.; Hueso,~L.~E. Activating the molecular spinterface.
  \emph{Nat Mater} \textbf{2017}, \emph{16}, 507--515\relax
\mciteBstWouldAddEndPuncttrue
\mciteSetBstMidEndSepPunct{\mcitedefaultmidpunct}
{\mcitedefaultendpunct}{\mcitedefaultseppunct}\relax
\EndOfBibitem
\bibitem[Friedrich \latin{et~al.}(2015)Friedrich, Caciuc, Kiselev, Atodiresei,
  and Bl\"ugel]{skyhook1}
Friedrich,~R.; Caciuc,~V.; Kiselev,~N.~S.; Atodiresei,~N.; Bl\"ugel,~S.
  Chemically functionalized magnetic exchange interactions of hybrid
  organic-ferromagnetic metal interfaces. \emph{Phys Rev B} \textbf{2015},
  \emph{91}, 115432\relax
\mciteBstWouldAddEndPuncttrue
\mciteSetBstMidEndSepPunct{\mcitedefaultmidpunct}
{\mcitedefaultendpunct}{\mcitedefaultseppunct}\relax
\EndOfBibitem
\bibitem[Callsen \latin{et~al.}(2013)Callsen, Caciuc, Kiselev, Atodiresei, and
  Bl\"ugel]{skyhooks1}
Callsen,~M.; Caciuc,~V.; Kiselev,~N.; Atodiresei,~N.; Bl\"ugel,~S. Magnetic
  Hardening Induced by Nonmagnetic Organic Molecules. \emph{Phys. Rev. Lett.}
  \textbf{2013}, \emph{111}, 106805\relax
\mciteBstWouldAddEndPuncttrue
\mciteSetBstMidEndSepPunct{\mcitedefaultmidpunct}
{\mcitedefaultendpunct}{\mcitedefaultseppunct}\relax
\EndOfBibitem
\bibitem[Brede \latin{et~al.}(2014)Brede, Atodiresei, Caciuc, Bazarnik,
  Al-Zubi, Bl\"ugel, and Wiesendanger]{skyhooks2}
Brede,~J.; Atodiresei,~N.; Caciuc,~V.; Bazarnik,~M.; Al-Zubi,~A.; Bl\"ugel,~S.;
  Wiesendanger,~R. Long-range magnetic coupling between nanoscale
  organic{\textendash}metal hybrids mediated by a nanoskyrmion lattice.
  \emph{Nat Nanotechnol} \textbf{2014}, \emph{9}, 1018--1023\relax
\mciteBstWouldAddEndPuncttrue
\mciteSetBstMidEndSepPunct{\mcitedefaultmidpunct}
{\mcitedefaultendpunct}{\mcitedefaultseppunct}\relax
\EndOfBibitem
\bibitem[Ormaza \latin{et~al.}(2016)Ormaza, Fern{\'{a}}ndez, Ilyn,
  Maga{\~{n}}a, Xu, Verstraete, Gastaldo, Valbuena, Gargiani, Mugarza, Ayuela,
  Vitali, Blanco-Rey, Schiller, and Ortega]{Ormaza.2016}
Ormaza,~M.; Fern{\'{a}}ndez,~L.; Ilyn,~M.; Maga{\~{n}}a,~A.; Xu,~B.;
  Verstraete,~M.~J.; Gastaldo,~M.; Valbuena,~M.~A.; Gargiani,~P.; Mugarza,~A.;
  Ayuela,~A.; Vitali,~L.; Blanco-Rey,~M.; Schiller,~F.; Ortega,~J.~E. High
  Temperature Ferromagnetism in a GdAg$_2$ Monolayer. \emph{Nano Lett}
  \textbf{2016}, \emph{16}, 4230--4235\relax
\mciteBstWouldAddEndPuncttrue
\mciteSetBstMidEndSepPunct{\mcitedefaultmidpunct}
{\mcitedefaultendpunct}{\mcitedefaultseppunct}\relax
\EndOfBibitem
\bibitem[Seidel \latin{et~al.}(2019)Seidel, Dup\'e, Mousavion, Walther,
  Medjanik, Vasilyev, Babenkov, Ellguth, Maniraj, Sinova, Sch\"onhense, Elmers,
  Stadtm\"uller, and Aeschlimann]{Seidel2019}
Seidel,~J.; Dup\'e,~B.; Mousavion,~S.; Walther,~E.~S.; Medjanik,~K.;
  Vasilyev,~D.; Babenkov,~S.; Ellguth,~M.; Maniraj,~M.; Sinova,~J.;
  Sch\"onhense,~G.; Elmers,~H.~J.; Stadtm\"uller,~B.; Aeschlimann,~M. Exchange
  Splitting of a Hybrid Surface State and Ferromagnetic Order in a 2D Surface
  Alloy, arXiv:1906.03780. 2019\relax
\mciteBstWouldAddEndPuncttrue
\mciteSetBstMidEndSepPunct{\mcitedefaultmidpunct}
{\mcitedefaultendpunct}{\mcitedefaultseppunct}\relax
\EndOfBibitem
\bibitem[Wagner and Muilenberg(1979)Wagner, and Muilenberg]{wagner1979handbook}
Wagner,~C.; Muilenberg,~G. \emph{Handbook of X-ray Photoelectron Spectroscopy:
  A Reference Book of Standard Data for Use in X-ray Photoelectron
  Spectroscopy}; Perkin-Elmer, 1979\relax
\mciteBstWouldAddEndPuncttrue
\mciteSetBstMidEndSepPunct{\mcitedefaultmidpunct}
{\mcitedefaultendpunct}{\mcitedefaultseppunct}\relax
\EndOfBibitem
\bibitem[Gierz \latin{et~al.}(2010)Gierz, Stadtm{\"u}ller, Vuorinen, Lindroos,
  Meier, Dil, Kern, and Ast]{Gierz.2010}
Gierz,~I.; Stadtm{\"u}ller,~B.; Vuorinen,~J.; Lindroos,~M.; Meier,~F.;
  Dil,~J.~H.; Kern,~K.; Ast,~C.~R. {Structural influence on the Rashba-type
  spin splitting in surface alloys}. \emph{{Phys Rev B}} \textbf{2010},
  \emph{81}, 245430\relax
\mciteBstWouldAddEndPuncttrue
\mciteSetBstMidEndSepPunct{\mcitedefaultmidpunct}
{\mcitedefaultendpunct}{\mcitedefaultseppunct}\relax
\EndOfBibitem
\bibitem[van Straaten \latin{et~al.}(2018)van Straaten, Franke, Bocquet, Tautz,
  and Kumpf]{VANSTRAATEN2018106}
van Straaten,~G.; Franke,~M.; Bocquet,~F.~C.; Tautz,~F.~S.; Kumpf,~C.
  Non-dipolar effects in photoelectron-based normal incidence X-ray standing
  wave experiments. \emph{J Electron Spectros Relat Phenomena} \textbf{2018},
  \emph{222}, 106 -- 116\relax
\mciteBstWouldAddEndPuncttrue
\mciteSetBstMidEndSepPunct{\mcitedefaultmidpunct}
{\mcitedefaultendpunct}{\mcitedefaultseppunct}\relax
\EndOfBibitem
\bibitem[Bocquet \latin{et~al.}(2019)Bocquet, Mercurio, Franke, van Straaten,
  Wei\ss, Soubatch, Kumpf, and Tautz]{toricelli}
Bocquet,~F.; Mercurio,~G.; Franke,~M.; van Straaten,~G.; Wei\ss,~S.;
  Soubatch,~S.; Kumpf,~C.; Tautz,~F. Torricelli: A software to determine atomic
  spatial distributions from normal incidence x-ray standing wave data.
  \emph{Comput Phys Commun} \textbf{2019}, \emph{235}, 502 -- 513\relax
\mciteBstWouldAddEndPuncttrue
\mciteSetBstMidEndSepPunct{\mcitedefaultmidpunct}
{\mcitedefaultendpunct}{\mcitedefaultseppunct}\relax
\EndOfBibitem
\bibitem[Hauschild \latin{et~al.}(2010)Hauschild, Temirov, Soubatch, Bauer,
  Sch\"{o}ll, Cowie, Lee, Tautz, and Sokolowski]{Hauschild2010}
Hauschild,~A.; Temirov,~R.; Soubatch,~S.; Bauer,~O.; Sch\"{o}ll,~A.; Cowie,~B.
  C.~C.; Lee,~T.-L.; Tautz,~F.~S.; Sokolowski,~M. Normal-incidence x-ray
  standing-wave determination of the adsorption geometry of {PTCDA} on Ag(111):
  Comparison of the ordered room-temperature and disordered low-temperature
  phases. \emph{Phys Rev B} \textbf{2010}, \emph{81}\relax
\mciteBstWouldAddEndPuncttrue
\mciteSetBstMidEndSepPunct{\mcitedefaultmidpunct}
{\mcitedefaultendpunct}{\mcitedefaultseppunct}\relax
\EndOfBibitem
\bibitem[Mercurio \latin{et~al.}(2013)Mercurio, Bauer, Willenbockel, Fairley,
  Reckien, Schmitz, Fiedler, Soubatch, Bredow, Sokolowski, and
  Tautz]{Mercurio2013}
Mercurio,~G.; Bauer,~O.; Willenbockel,~M.; Fairley,~N.; Reckien,~W.;
  Schmitz,~C.~H.; Fiedler,~B.; Soubatch,~S.; Bredow,~T.; Sokolowski,~M.;
  Tautz,~F.~S. Adsorption height determination of nonequivalent C and O species
  of {PTCDA} on Ag(110) using x-ray standing waves. \emph{Phys Rev B}
  \textbf{2013}, \emph{87}\relax
\mciteBstWouldAddEndPuncttrue
\mciteSetBstMidEndSepPunct{\mcitedefaultmidpunct}
{\mcitedefaultendpunct}{\mcitedefaultseppunct}\relax
\EndOfBibitem
\bibitem[Tautz(2007)]{Tautz2007}
Tautz,~F. Structure and bonding of large aromatic molecules on noble metal
  surfaces: The example of {PTCDA}. \emph{Prog Surf Sci} \textbf{2007},
  \emph{82}, 479--520\relax
\mciteBstWouldAddEndPuncttrue
\mciteSetBstMidEndSepPunct{\mcitedefaultmidpunct}
{\mcitedefaultendpunct}{\mcitedefaultseppunct}\relax
\EndOfBibitem
\bibitem[Kawabe \latin{et~al.}(2008)Kawabe, Yamane, Sumii, Koizumi, Ouchi,
  Seki, and Kanai]{Kawabe2008}
Kawabe,~E.; Yamane,~H.; Sumii,~R.; Koizumi,~K.; Ouchi,~Y.; Seki,~K.; Kanai,~K.
  A role of metal d-band in the interfacial electronic structure at
  organic/metal interface: {PTCDA} on Au, Ag and Cu. \emph{Org Electron}
  \textbf{2008}, \emph{9}, 783--789\relax
\mciteBstWouldAddEndPuncttrue
\mciteSetBstMidEndSepPunct{\mcitedefaultmidpunct}
{\mcitedefaultendpunct}{\mcitedefaultseppunct}\relax
\EndOfBibitem
\bibitem[Romaner \latin{et~al.}(2009)Romaner, Nabok, Puschnig, Zojer, and
  Ambrosch-Draxl]{Romaner2009}
Romaner,~L.; Nabok,~D.; Puschnig,~P.; Zojer,~E.; Ambrosch-Draxl,~C. Theoretical
  study of {PTCDA} adsorbed on the coinage metal surfaces, Ag(111), Au(111) and
  Cu(111). \emph{New J Phys} \textbf{2009}, \emph{11}, 053010\relax
\mciteBstWouldAddEndPuncttrue
\mciteSetBstMidEndSepPunct{\mcitedefaultmidpunct}
{\mcitedefaultendpunct}{\mcitedefaultseppunct}\relax
\EndOfBibitem
\bibitem[Cottin \latin{et~al.}(2014)Cottin, Lobo-Checa, Schaffert, Bobisch,
  M\"{o}ller, Ortega, and Walter]{Cottin2014}
Cottin,~M.~C.; Lobo-Checa,~J.; Schaffert,~J.; Bobisch,~C.~A.; M\"{o}ller,~R.;
  Ortega,~J.~E.; Walter,~A.~L. A chemically inert Rashba split interface
  electronic structure of C60, {FeOEP} and {PTCDA} on {BiAg}2/Ag(111)
  substrates. \emph{New J Phys} \textbf{2014}, \emph{16}, 045002\relax
\mciteBstWouldAddEndPuncttrue
\mciteSetBstMidEndSepPunct{\mcitedefaultmidpunct}
{\mcitedefaultendpunct}{\mcitedefaultseppunct}\relax
\EndOfBibitem
\bibitem[Seidel \latin{et~al.}(2020)Seidel, Kelly, Franke, Kumpf, Cinchetti,
  Aeschlimann, and Stadtm\"uller]{SeidelSnAg2019}
Seidel,~J.; Kelly,~L.~L.; Franke,~M.; Kumpf,~C.; Cinchetti,~M.;
  Aeschlimann,~M.; Stadtm\"uller,~B. Vertical bonding distances and interfacial
  band structure of PTCDA on a Sn-Ag surface alloy, arXiv:2002.01831.
  2020\relax
\mciteBstWouldAddEndPuncttrue
\mciteSetBstMidEndSepPunct{\mcitedefaultmidpunct}
{\mcitedefaultendpunct}{\mcitedefaultseppunct}\relax
\EndOfBibitem
\bibitem[Tiba \latin{et~al.}(2006)Tiba, de~Jonge, Koopmans, and
  Jonkman]{Tiba2006}
Tiba,~M.~V.; de~Jonge,~W. J.~M.; Koopmans,~B.; Jonkman,~H.~T. Morphology and
  electronic properties of the pentacene on cobalt interface. \emph{J Appl
  Phys} \textbf{2006}, \emph{100}, 093707\relax
\mciteBstWouldAddEndPuncttrue
\mciteSetBstMidEndSepPunct{\mcitedefaultmidpunct}
{\mcitedefaultendpunct}{\mcitedefaultseppunct}\relax
\EndOfBibitem
\bibitem[Droghetti \latin{et~al.}(2016)Droghetti, Thielen, Rungger, Haag,
  Grossmann, St\"ockl, Stadtm\"uller, Aeschlimann, Sanvito, and
  Cinchetti]{Droghetti.2016}
Droghetti,~A.; Thielen,~P.; Rungger,~I.; Haag,~N.; Grossmann,~N.; St\"ockl,~J.;
  Stadtm\"uller,~B.; Aeschlimann,~M.; Sanvito,~S.; Cinchetti,~M. {Dynamic spin
  filtering at the Co/Alq3 interface mediated by weakly coupled second layer
  molecules}. \emph{{Nat Commun}} \textbf{2016}, \emph{7}, 12668\relax
\mciteBstWouldAddEndPuncttrue
\mciteSetBstMidEndSepPunct{\mcitedefaultmidpunct}
{\mcitedefaultendpunct}{\mcitedefaultseppunct}\relax
\EndOfBibitem
\bibitem[Mirhosseini \latin{et~al.}(2010)Mirhosseini, Ernst, Ostanin, and
  Henk]{Mirhosseini2010}
Mirhosseini,~H.; Ernst,~A.; Ostanin,~S.; Henk,~J. Tuning independently the
  Fermi energy and spin splitting in Rashba systems: ternary surface alloys on
  Ag(111). \emph{Journal of Physics: Condensed Matter} \textbf{2010},
  \emph{22}, 385501\relax
\mciteBstWouldAddEndPuncttrue
\mciteSetBstMidEndSepPunct{\mcitedefaultmidpunct}
{\mcitedefaultendpunct}{\mcitedefaultseppunct}\relax
\EndOfBibitem
\bibitem[Meier \latin{et~al.}(2009)Meier, Petrov, Guerrero, Mudry, Patthey,
  Osterwalder, and Dil]{Meier2009}
Meier,~F.; Petrov,~V.; Guerrero,~S.; Mudry,~C.; Patthey,~L.; Osterwalder,~J.;
  Dil,~J.~H. Unconventional Fermi surface spin textures in
  {theBixPb}1-x/Ag(111)surface alloy. \emph{Phys Rev B} \textbf{2009},
  \emph{79}\relax
\mciteBstWouldAddEndPuncttrue
\mciteSetBstMidEndSepPunct{\mcitedefaultmidpunct}
{\mcitedefaultendpunct}{\mcitedefaultseppunct}\relax
\EndOfBibitem
\bibitem[Pacil{\'{e}} \latin{et~al.}(2006)Pacil{\'{e}}, Ast, Papagno, Silva,
  Moreschini, Falub, Seitsonen, and Grioni]{Pacil2006}
Pacil{\'{e}},~D.; Ast,~C.~R.; Papagno,~M.; Silva,~C.~D.; Moreschini,~L.;
  Falub,~M.; Seitsonen,~A.~P.; Grioni,~M. Electronic structure of an
  {orderedPb}-Ag(111)surface alloy: Theory and experiment. \emph{Phys Rev B}
  \textbf{2006}, \emph{73}\relax
\mciteBstWouldAddEndPuncttrue
\mciteSetBstMidEndSepPunct{\mcitedefaultmidpunct}
{\mcitedefaultendpunct}{\mcitedefaultseppunct}\relax
\EndOfBibitem
\bibitem[Thole \latin{et~al.}(1985)Thole, van~der Laan, and
  Sawatzky]{Thole1985}
Thole,~B.~T.; van~der Laan,~G.; Sawatzky,~G.~A. Strong Magnetic Dichroism
  Predicted in {theM}4, 5X-Ray Absorption Spectra of Magnetic Rare-Earth
  Materials. \emph{Phys. Rev. Lett.} \textbf{1985}, \emph{55}, 2086--2088\relax
\mciteBstWouldAddEndPuncttrue
\mciteSetBstMidEndSepPunct{\mcitedefaultmidpunct}
{\mcitedefaultendpunct}{\mcitedefaultseppunct}\relax
\EndOfBibitem
\end{mcitethebibliography}
\end{document}


\title{Modification of the charge and magnetic order of a low dimensional ferromagnet by molecule-surface bonding\\Supplemental Material}

\author{Johannes Seidel}
\email[]{jseidel@rhrk.uni-kl.de}
\affiliation{Department of Physics and Research Center OPTIMAS, University of Kaiserslautern, Erwin-Schroedinger-Strasse 46, 67663 Kaiserslautern, Germany}
\author{Eva S. Walther}
\affiliation{Department of Physics and Research Center OPTIMAS, University of Kaiserslautern, Erwin-Schroedinger-Strasse 46, 67663 Kaiserslautern, Germany}
\author{Sina Mousavion}
\affiliation{Department of Physics and Research Center OPTIMAS, University of Kaiserslautern, Erwin-Schroedinger-Strasse 46, 67663 Kaiserslautern, Germany}

\author{Dominik Jungkenn}
\affiliation{Department of Physics and Research Center OPTIMAS, University of Kaiserslautern, Erwin-Schroedinger-Strasse 46, 67663 Kaiserslautern, Germany}
\author{Markus Franke}
\affiliation{Peter Gr\"unberg Institut (PGI-3), Forschungszentrum J\"ulich, 52425 J\"ulich, Germany}
\affiliation{J\"ulich-Aachen Research Alliance (JARA) -- Fundamentals of Future Information Technology, 52425 J\"ulich, Germany}
\author{Leah L. Kelly}
\affiliation{Department of Physics and Research Center OPTIMAS, University of Kaiserslautern, Erwin-Schroedinger-Strasse 46, 67663 Kaiserslautern, Germany}
\author{Ahmed Alhassanat}
\affiliation{Johannes-Gutenberg-Universit\"at Mainz, Institut f\"ur Physik, 55128 Mainz, Germany}
\author{Hans-Joachim Elmers}
\affiliation{Johannes-Gutenberg-Universit\"at Mainz, Institut f\"ur Physik, 55128 Mainz, Germany}
\affiliation{Graduate School of Excellence Materials Science in Mainz, Erwin Schroedinger Strasse 46, 67663 Kaiserslautern, Germany}
\author{Mirko Cinchetti}
\affiliation{Experimentelle Physik VI, Technische Universit\"at Dortmund, 44221 Dortmund, Germany}
\author{Christian Kumpf}
\affiliation{Peter Gr\"unberg Institut (PGI-3), Forschungszentrum J\"ulich, 52425 J\"ulich, Germany}
\author{Martin Aeschlimann}
\affiliation{Department of Physics and Research Center OPTIMAS, University of Kaiserslautern, Erwin-Schroedinger-Strasse 46, 67663 Kaiserslautern, Germany}
\author{Benjamin Stadtm\"uller}
\affiliation{Department of Physics and Research Center OPTIMAS, University of Kaiserslautern, Erwin-Schroedinger-Strasse 46, 67663 Kaiserslautern, Germany}
\affiliation{Graduate School of Excellence Materials Science in Mainz, Erwin Schroedinger Strasse 46, 67663 Kaiserslautern, Germany}

\date{\today}

\maketitle

\section{Experimental details}

\subsection{Sample preparation}
All experiments have been carried out at a base pressure of below 5*10$^{-10}\,$mbar. The Ag(111) single crystals have been cleaned by multiple Ar Ion sputtering cycles of various energies (0.5 - 2.0 keV, I$_{drain}=8\,\mu A$) and subsequent annealing cycles of 20 minutes up to 730 K. Sample cleanness was confirmed by the existence and lineshape of the Shockley surface state. XPS was checked for typical contamination species like carbon or oxygen. Cleanness was confirmed by absence of those. The dysprosium surface alloy is fabricated using a commercial Focus EFM3 evaporator. The surface alloy is formed at elevated sample temperatures (T=600 K). The sample quality is confirmed by presence and lineshape of the superstructure spots in LEED and the lineshape of the Dy 3d core level. The organic molecules are deposited by a commercial kentax evaporator with a cell temperature of T=560 K (PTCDA). The flux was checked by mass spectroscopy (one monolayer in approximately 10 minutes) and checking C1s XPS peak areas to confirm the coverage on the sample itself. 
\subsection{Experimental methods}
The UPS data has been accessed by momentum microscopy at the NanoESCA endstation of the Elettra synchrotron Triest. This endstation is equipped with a double hemispherical analyser and is optimized to image the momentum distribution of photoelectrons. The XMCD measurements were carried out at the VEKMAG instrument at Helmholtz-Zentrum Berlin. 
The core level and normal incidence standing waves experiments were conducted at the hard x-ray photoemission (HAXPES) and x-ray standing wave (XSW) end station at beamline I09 at the Diamond  Light Source synchrotron in Didcot, UK\cite{Lee.2018}. This end station is equipped with a hemispherical electron analyzer (Scienta EW4000 HAXPES), which is mounted perpendicular to the direction of the incoming photon beam. This analyzer has an angular acceptance of $\pm$ 30$^{\circ}$ and an energy resolution of $150\,$meV using the analyzer settings of our experiments ($E_{pass}$=$100\,$eV, $d_{slit}=0.5\,$mm).
The NIXSW method is a powerful tool to investigate the adsorption height of chemically inequivalent atoms above a single crystalline surface with a precision of better than $\Delta z \leq 0.04\,$\AA. A detailed description of the method can be found in \cite{WOODRUFF19981,Woodruff_2005,ZEGENHAGEN1993202}. Here, we will just give a short introduction into the basic principle:
\\
Photon energies fulfilling the Bragg condition ($H=k_H-k_0$) of a specific reflection ($H=(h k l )$) lead to the formation of a standing wave field above the single crystal surface due to the interference between the incoming and back-reflected x-ray wave fields. Scanning the photon energy in a range of several eV around the Bragg condition leads to a phase shift of the relative complex amplitudes between the incoming and back-reflected waves by $\pi$. This phase shift moves the maxima of the the standing wave field incrementally by half a lattice spacing $d_{(h k l)}$ perpendicular to the Bragg planes. The shift of the standing wave field leads to a change of the photon density at any specific position $z$ above the single crystal surface as a function of the photon energy. An atom at a position $z$ above the single crystal shows a varying XPS yield depending on the position of the standing wave field, e.g. the photon energy used.
 The experimentally observed yield curve $I(E)\,$ can be described by:

\begin{equation}
\begin{split}
I(E)=1+S_R R(E)+2\vert S_I \vert \sqrt{R(E)} \\\
+F^H \cos{(\nu(E)-2\pi P^H+\Psi)},
\end{split}
\label{Eq:xsw}
\end{equation}
where $R(E)$ is the x-ray reflectivity of the Bragg reflection with its complex amplitude $\sqrt{R(E)}\,$ and phase $\nu (E)$. $S_R$, $\vert S_I \vert\,$and $\Psi\,$ are correction parameters for non-dipolar contributions to the photoemission yield. For more details, see\cite{WOODRUFF2005187,SCHREIBER2001L519,VARTANYANTS2005196,VANSTRAATEN2018106}.
 The fitting parameters of the NIXSW analysis are the coherent fraction $F^H$ and coherent position $P^H$. The coherent position $P^H$ can be interpreted as an average height of the chemical species with respect to the Bragg planes: $z=d(h\, k\, l)\times (P^H+n)$, n= 0, 1, 2,\dots , where n is the number of Bragg planes located between the surface plane and the average position of the species. The coherent fraction $F^H$ is a parameter quantifying the vertical order of this atomic species. Perfect vertical order corresponds to a coherent fraction $F^H=1$, while a homogeneous distribution of adsorbates between two Bragg planes would lead to a coherent fraction $F^H=0$. Single layers of molecular adsorbates are reported to show coherent fractions of carbon, nitrogen or oxygen in the range of $F^H=0.8$ for highly ordered films\cite{Kroger_2010}. This reduction is attributed to the molecular vibrations and small structural defects of the non-perfect molecular film which can significantly influence the vertical order of the different chemical species.
  \section{Lateral ordering}
\begin{figure}
\includegraphics[width=15 cm]{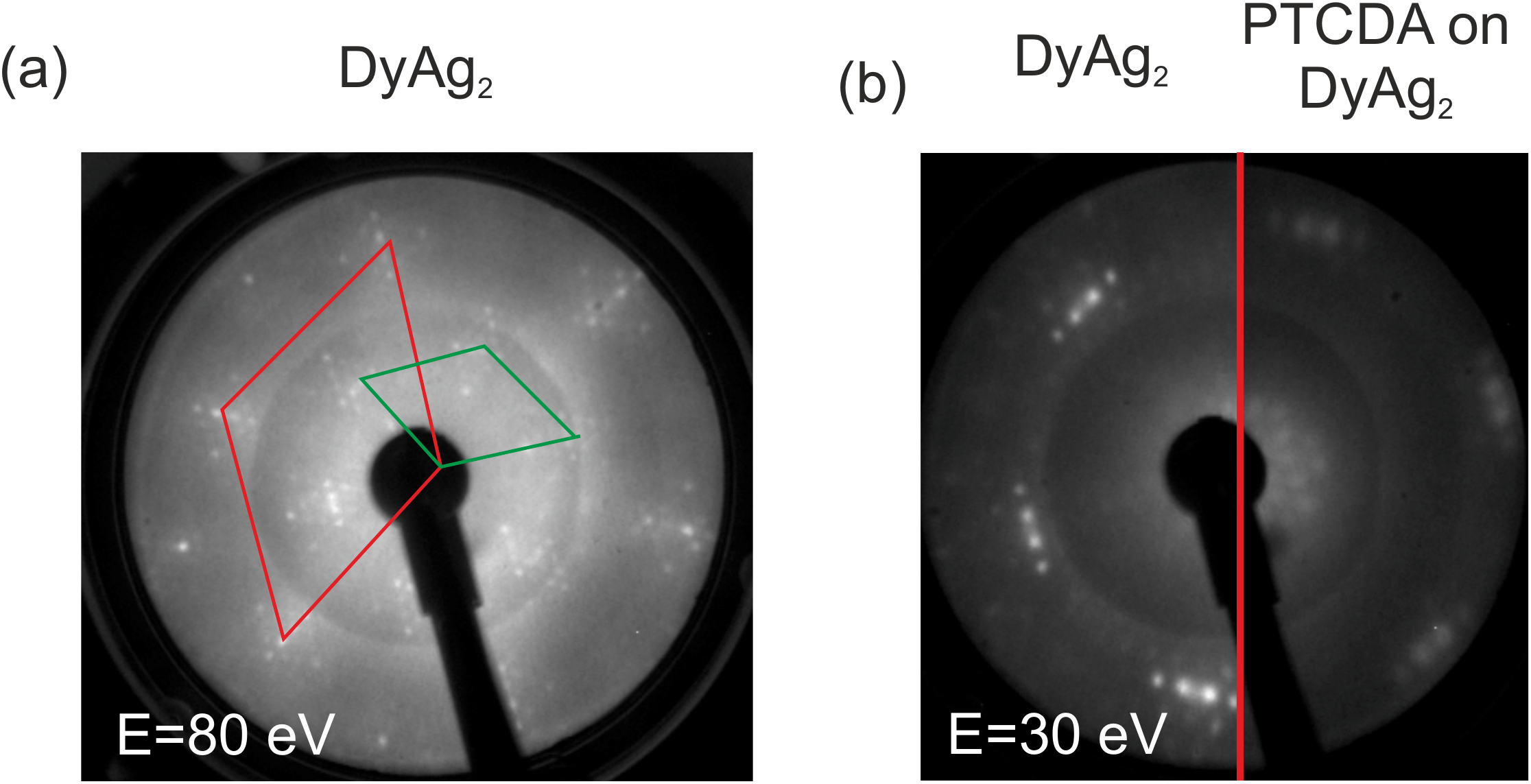}
\caption{LEED images for the discussed sample systems. (a) Clean DyAg$_2$. Ag(111) unit cell sketched in red, superstructure unit cell sketched in green. (b)left half: clean DyAg$_2$, right half: monolayer PTCDA adsorbed on DyAg$_2$.}
\label{Fig:Fig1}
\end{figure}
The lateral structure of the surface alloy is in full agreement with a local $\sqrt{3}\times \sqrt{3}R30\,$ superstructure. Due to a lattice missmatch between the DyAg$_2$ surface alloy and the underlying Ag single crystal an additional Moir\'e pattern of $16 \times 16\,$ is observed. We show an exemplary LEED image for the clean surface alloy in Fig.~\ref{Fig:Fig1}a. We sketch the Ag(111) unit cell as red rhomb, the DyAg$_2$ unit cell in green. For adsorption of PTCDA (Fig.~\ref{Fig:Fig1}(b)) we observe no clear molecular spots, only a dampening of the substrate intensity. These observations indicate a completely disordered molecular film.

\section{UPS - PTCDA and electronic substrate structure}
\begin{figure}
\includegraphics[width=15 cm]{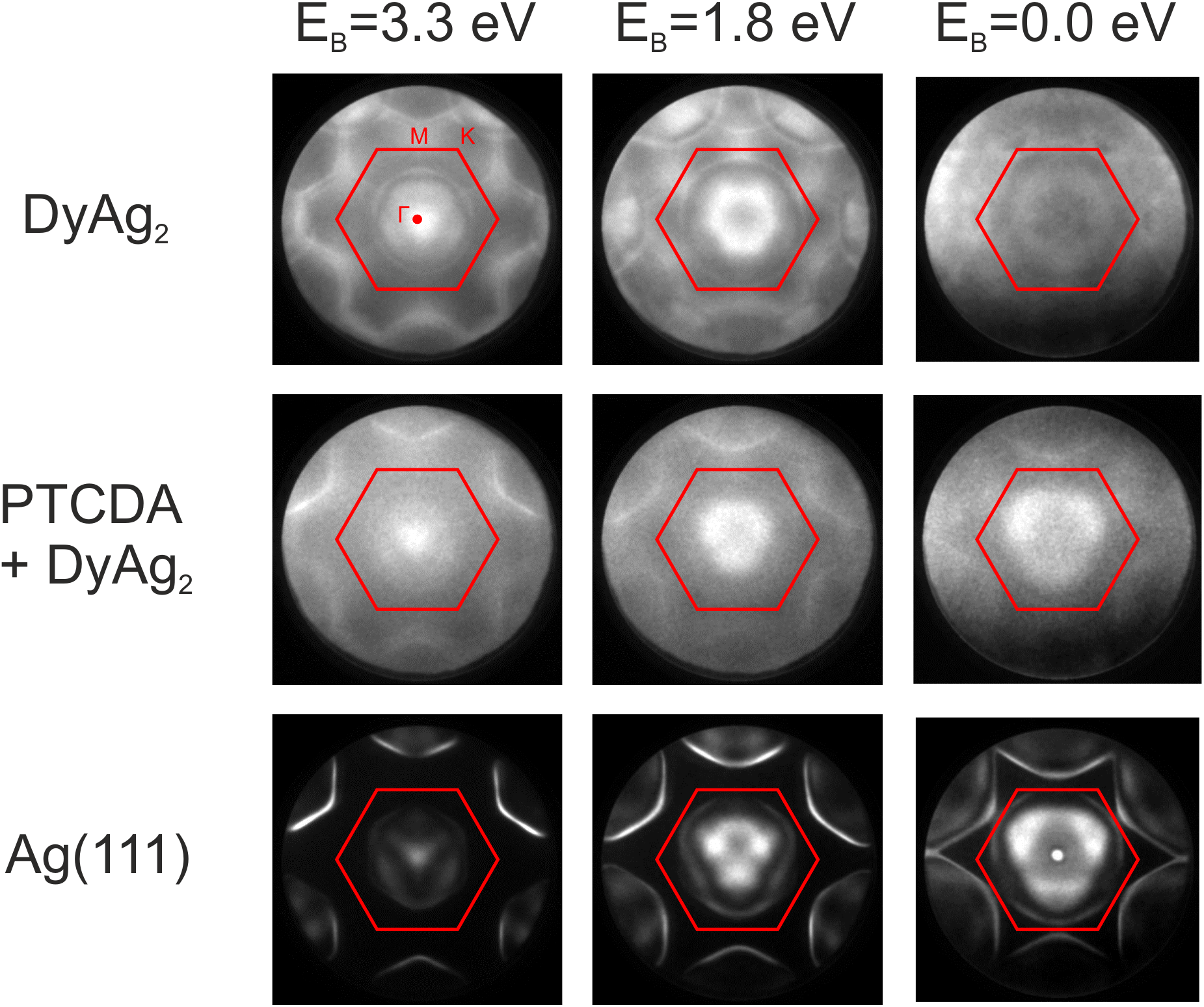}
\caption{Constant maps for three different binding energies in the valence band region. Top line clean DyAg$_2$. Middle line: monolayer of PTCDA adsorbed on DyAg$_2$. Bottom line: bare Ag(111) emission. SBZ of Ag(111) is sketched in red.}
\label{Fig:Fig3}
\end{figure}

We want to illustrate the change of electronic substrate structure for adsorption of PTCDA on DyAg$_2$ with several constant energy maps. We present three binding energies in Fig.~\ref{Fig:Fig3} for the clean surface alloy (top row), PTCDA/DyAg$_2$ (middle row) and Ag(111) (bottom row). Note that for adsorption of PTCDA not only the center emission pattern is changed but also the ring-like feature at the K-point at E$_F$ disappears. In addition, the features at the M point for E$_B$=1.8 eV vanish as well. These findings point to a significant change of the electronic structure of the DyAg$_2$ surface alloy upon adsorption of PTCDA.
